# Structural, Chemical and Optical Properties of Cerium Dioxide Film Prepared by Atomic Layer Deposition on TiN and Si Substrates

*S. Vangelista[1], R. Piagge[2], S. Ek[3], T. Sarnet[3], G. Ghidini[2], C. Martella[1] and A. Lamperti[1]*

[1]CNR-IMM - MDM Laboratory, Via C. Olivetti 2, Agrate Brianza (MB) I-20864 Italy

[2]STMicroelectronics, Via C. Olivetti 2, Agrate Brianza (MB) I-20864 Italy

[3]Picosun Oy, Tietotie 3, Espoo FI-02150 Finland

Thin films of cerium dioxide ($CeO_2$) were deposited by atomic layer deposition (ALD) at 250 °C on both Si and TiN substrates. The ALD growth produces $CeO_2$ films with polycrystalline cubic phase on both substrates. However, the films show a preferential orientation along <200> crystallographic direction for $CeO_2$/Si or <111> for $CeO_2$/TiN, as revealed by X-ray diffraction. Additionally, $CeO_2$ films differ in interface roughness depending on the substrate. Furthermore, the relative concentration of $Ce^{3+}$ is 22.0% in $CeO_2$/Si and around 18% in $CeO_2$/TiN, as obtained by X-ray photoelectron spectroscopy (XPS). Such values indicate a ~10% off-stoichiometry and are indicative of the presence of oxygen vacancies in the films. Nonetheless, $CeO_2$ bandgap energy and refractive index at 550 nm are 3.54±0.63 eV and 2.3 for $CeO_2$/Si, and 3.63±0.18 eV and 2.4 for $CeO_2$/TiN, respectively. Our results extend the knowledge on the structural and chemical properties of ALD-deposited $CeO_2$ either on Si or TiN substrates, underlying films differences and similarities, thus contributing to boost the use of $CeO_2$ through ALD deposition as foreseen in a wide number of applications.

1. INTRODUCTION

In the last years, $CeO_2$ based materials have attracted much attention due to their wide use in many application areas such as catalysis, hydrogen production, gas sensing and electrodes in fuel cells [1,2 ,3 ,4 ,5]. In microelectronics, $CeO_2$ has been considered as high κ-gate oxide material due to its unique properties: moderate band gap (3–3.6 eV), high dielectric constant (κ: 23–26), high refractive index (n: 2.2–2.8) and high dielectric strength (~2.6 MV cm$^{-1}$) [6]. $CeO_2$ is also suitable for Si based metal oxide semiconductor (MOS) devices due to its small lattice mismatch (−0.35%) with Si that favors its epitaxial growth on different silicon surfaces [7] and low interface-state density (~ $10^{11}$ cm$^{-2}$eV$^{-1}$) [5,8]. $CeO_2$ use was explored also for non-volatile memory devices in metal-insulator-metal stacks [9]. For catalytic applications $CeO_2$ has to be grown on metallic substrates; however, few studies exist on cerium oxide film grown on metals and with ultra-low thickness (mono or few layers) aiming at depicting the epitaxial growth relation [10and references therein].

$CeO_2$ deposition is achieved by using a variety of growth techniques, such as e-beam [11], physical vapor deposition [12], RF-magnetron sputtering [13], chemical vapor deposition (CVD) [14,15 ,16 ,17 ,18], even in sub-stoichiometric form ($CeO_{2-x}$, 0<x<0.4) [19]. However, when highly stoichiometric oxide is required, atomic layer deposition (ALD) is the most suitable technique. ALD is a self-limiting technique, based on sequential surface reactions. Moreover, ALD allows the atomic layer control of the film thickness with large-area uniformity and conformality also on 3D surfaces. Further, thin films deposition can be performed at lower temperatures than other vacuum deposition techniques such as pulsed laser deposition or CVD, a process condition that guarantees low thermal budget, no or limited interdiffusion phenomena, and the possibility to use temperature-sensitive substrates. ALD deposition of $CeO_2$ on Si or Si/$SiO_2$ substrates has been explored by using various precursors [20,21 ,22 ,23 ,24 ,25 ,26 ,27], obtaining as-deposited film with polycrystalline structure [22-25] which influences the dielectric behavior [24]. However, ALD of $CeO_2$ on metal substrates (or electrically behaving as metals, such as TiN) has not been explored yet.

In this work, we aim to a deep understanding of the structural properties of ALD-deposited $CeO_2$ either on Si or TiN substrate, underlying the differences and similarities in the properties of the obtained films. Our results will contribute to give the basis for the use of $CeO_2$ through ALD deposition as foreseen in a wide number of applications, in particular in microelectronics and catalysis.

## 2. Experimental section

Cerium dioxide thin films were deposited in a Picosun R-200 Advanced ALD reactor by Picosun Oy. The precursor $Ce(thd)_4$ (thd = 2,2,6,6-tetramethyl-3,5-heptanedione) was evaporated at the temperature of 140 °C and ozone has been used as oxidizing agent (concentration 19%). Both precursors were transported into the reactor chamber by $N_2$ carrier gas flow. Based on the results reported in ref. [22], the deposition temperature was chosen as 250 °C, the pulse times for $Ce(thd)_4$ and ozone were 1.0 s and 2.5 s respectively, and the purge time with nitrogen after pulses was 1.5 s. Under these conditions the growth rate was 0.32 Å/cycle. The targeted oxide thickness on both substrates was 25 nm. The deposition was done on both Si and TiN-coated 7x7 $cm^2$ Si substrates simultaneously, by placing them on an 8" silicon wafer which was put in the wafer holder. TiN coating layers, nominally 7 nm thick, have been deposited by CVD.

X-ray reflectivity (XRR) and diffraction at grazing incidence (GIXRD) have been performed using an XRD3000 diffractometer (Italstructure) with monochromated X-ray Cu Kα radiation (wavelength 0.154 nm) and beam size defined by slits aperture of 0.1 × 6 mm. Data fitting, performed using MAUD program [28], allowed us to determine thickness, interface and surface roughness, and electronic density. MAUD has also been used for Rietveld refinement of GIXRD patterns, allowing to determine $CeO_2$ cell parameter and grain size [29].

The morphology of the samples was investigated in tapping mode by means of an AFM-Bruker commercial system, equipped with ultrasharp silicon probes (nominal tip radius ~10 nm). Root-mean-square (rms) roughness and other statistical parameters of the surface morphology were quantitatively derived from the topographies employing the free-available WSxM software.

Focus-ion-beam scanning electron microscopy (SEM) has been used to prepare cross sections of selected samples. Energy Dispersive X-ray Analysis (EDX, Oxford X-MAS 80 $mm^2$) in STEM mode was acquired in a Transmission Electron Microscope (TEM, Fei Tecnai G2) operating at 200 kV. EDX allows for the identification of the chemical composition with a detection limit ~ 1% for the analysed elements.

The compositional depth profile of the $CeO_2$/TiN/Si and $CeO_2$/Si structures was also investigated by Time of Flight-Secondary Ion Mass Spectrometry (ToF-SIMS) by means of a $Cs^+$ ion beam (energy of 0.5 keV, ion current 38.5 nA) sputtering a 200 μm × 200 μm area, and $Ga^+$ ion beam (25 keV, 2.7 pA) for analysis over a 50 μm × 50 μm area centred on the sputtered crater and therein collecting secondary negative ions [30].

To gain a deep knowledge on the chemical state of the $CeO_2$ [31] films, X-ray photoelectron spectroscopy (XPS) measurements were performed on a PHI 5600 instrument (Physics Electronics Inc.) equipped with a monochromatic AlKα X-ray source (energy = 1486.6 eV) and a concentric hemispherical analyser. The spectra were collected at a take-off angle of 45° and band-pass energy of 23.50 eV. The instrument resolution is 0.5 eV. The spectra were aligned using *C 1s* peak (284.6 eV) as reference [32]. The experimental data were fitted with Gaussian-Lorentzian peaks and Shirley background by using XPSPeak program (version 4.1, Freeware, University of Warwick, United Kingdom). The spin-orbit splitting and doublet intensity ratio values have been fixed from literature [33].

Spectroscopic Ellipsometry (SE, Woollam M-2000 F) was used to acquire spectra in the visible-UV range from 1.24 to 5.05 eV. The ellipsometric angle ψ and phase difference Δ were recorded at an incidence angle of 75°. The optical dielectric function of $CeO_2$ was obtained by using a model of two Tauc–Lorentz (2TL) oscillators by using EASE program (J.A. Woollam Co., Inc.), allowing to obtain the refractive index and the direct and indirect band gaps.

### 3. Results and Discussion

XRR of as-deposited films on Si and TiN substrates are shown in **figure 1(a)** and in **figure 1(b)**, **respectively**. The values of thickness, roughness and electronic density obtained from the fitting are reported in **Table 1**. $CeO_2$ thickness is close to the target (25 nm) in both cases, as confirmed also by TEM (not shown). The interface roughness is higher in the $CeO_2$/TiN sample than in the $CeO_2$/Si sample; in addition, a very thin oxidation of TiN surface is detected. The calculated electronic density of $CeO_2$ films is slightly below the bulk value (1.79 $e^-/Å^3$), nonetheless in line with the considerations that we are measuring thin films at the nanoscale with the presence of contaminants (mainly C and OH) from $Ce(thd)_4$ precursor decomposition [22]. The surface roughness is limited to 1.4 nm. On the top of the as-deposited films the absorption of humidity is detected, a well-known phenomenon for $CeO_2$ [34], due to its intrinsic hygroscopicity. The presence of humidity at the surface has been detected also by performing AFM measurements of the samples, in particular for $CeO_2$/TiN. Despite the strong water-tip interaction, we managed to acquire the surface images of the samples (see the inset of the figure 1(a) and 1(b)) and to calculate grain size and RMS roughness: in both samples, we found around 15 nm and 1.2 nm, respectively. Thus, AFM analysis confirms XRR findings.

The crystallinity of the films was examined by GIXRD at fixed incidence angle of 2°. **Figure 2** shows the GIXRD patterns of as-deposited $CeO_2$ film on Si and TiN substrates. Both films are polycrystalline, exhibiting the typical peaks of a face centered cubic (fcc) structure (*Fm-3m* in space group notation), as reported in the crystallographic database [35]. No signal coming from the TiN layer is detected (TiN layer is 7.2 nm thick from XRR), indicating its amorphous nature – as confirmed by TEM (not shown).

More in detail, the observed relative intensity of $CeO_2$ peaks gives an indication of some preferential orientation. This evidence becomes clear by performing Rietveld refinement of the GIXRD data assuming the powder pattern as initial input (see the continuous lines on top of the experimental dot points in **figure 2**). For $CeO_2$ deposited on TiN (lower black dots with red line) the $I_{111}/I_{200}$ intensity ratio is higher than that of the powder spectrum, while for the $CeO_2$ deposited on Si (upper blue dots with light blue line) the $I_{111}/I_{200}$ intensity ratio is lower than that of the powder diffraction pattern, the latter observation in accordance with ref. [36].

These findings suggest that $CeO_2$ growth is influenced by the different nature of the surface/layer onto which it is deposited. It is likely that the presence of a higher roughness at the $CeO_2$/TiN interface with respect to the case of $CeO_2$/$SiO_2$ could contribute in changing the energy requirement for the nucleation of crystals with *(111)* orientation. Atomistic simulations [37,38] show that the formation of crystallites exposing *(100)* surfaces is unstable because of the dipole generated perpendicular to it. However, the formation of polycrystalline $CeO_2$ with all crystalline orientations is well-demonstrated experimentally by using different growth methods [22], [25], [39] and more generally it can be justified considering defects or charge compensating species, i.e. oxygen vacancies due to $Ce^{3+}$: $Ce^{4+}$ substitution. In our case, the contaminants (C and OH) from the precursor can act as source of defects, or oxygen vacancies could be present due to a different reactivity of TiN versus Si. Further, using Sherrer's formula applied to the *(111)* and *(200)* peaks to estimate the size of the crystallites, the grains with *(200)* orientation result to be slightly larger than the *(111)*-oriented grains for $CeO_2$ films grown on both TiN and Si substrates, as reported on **Table 1**. However, before any other consideration, we remind the isotropic approximation in the Sherrer's implies that the crystalline grains are considered spherical, i.e. with the same dimension in all the crystallographic directions. This is likely not to be realistic in thin films, and the grains belonging to different orientations assume different shape and/or different distribution through the layer thickness.

In order to understand the presence and the nature of the contaminants in $CeO_2$ films we performed EDX and ToF-SIMS analysis. In **figures 3(a)** and **3(b)** we report ToF-SIMS chemical depth profiles for $CeO_2$ deposited on Si and on TiN respectively, together with the corresponding chemical maps of the elements in the layers from EDX (upper panels). At first, we observe that the $CeO_2$/Si interface is very sharp (**figure 3(a)**), with the existence of a thin silicon native oxide layer. On the other side, the $CeO_2$/TiN interface is slight broad (**figure 3(b)**), with the signals related with TiN showing not negligible tails in the oxide film region. This is possibly due to the roughness of the $CeO_2$/TiN interface rather than an onset of Ti diffusion into the $CeO_2$ film, also considering the low growth temperature (250 °C). Moreover, ToF-SIMS allows following the signals coming from C and OH, which are the typical spurious elements incorporated during an ALD deposition when heptanedionate (thd) and ozone precursors are used [22]. By comparing the levels of the signals, we observe that C intensity is comparatively lower in $CeO_2$/TiN than in $CeO_2$/Si, while OH shows almost the same intensity. However, also in the worst case (i.e. $CeO_2$/Si) the correspondent C amount is only few atomic percent, as expected [22]. In any case, it is important to underline the presence of carbon, eventually in presence also of shortage of oxygen, since it can induce the reduction of $Ce^{4+}$ into $Ce^{3+}$, which is fundamental in the use of ceria as catalytic converter [40].

The presence of $Ce^{3+}$ can be related to defects in $CeO_2$, such as oxygen vacancies related with the oxygen storage capacity of $CeO_2$ [41], i.e. to a highly non-stoichiometric $CeO_{2-x}$ phase [42]. A $Ce_2O_3$-like phase cannot be in principle excluded, but we do not have indication of $Ce_2O_3$ phase in the diffraction profiles, even in films annealed at high temperature.

The off-stoichiometry of the oxide can be estimated from the XPS analysis. Despite the technique probing volume is limited to the outermost part of the $CeO_2$ layers, which can convey of the interaction with the atmosphere, from the $Ce^{3+}/Ce^{4+}$ ratio estimation we can have indication of the chemical stoichiometry of the whole oxide layer. **Figures 4(a)** and **4(b)** show the *Ce 3d* and *O 1s* spectra of $CeO_2$ on Si (upper panel) and on TiN (lower panel). The figures show also the fit obtained from the deconvolution of the spectra into the different components, similarly to that reported in refs. [18, 42, 43, 44]. The analysis of the *Ce 3d* spectrum is particularly complex since the deconvolution considers 5 different doublets related to the two oxidation states due to primary photoionization and shake-down satellites: three related to the $Ce^{4+}$ doublets and the other two related to the $Ce^{3+}$ doublets. These components are reported in **Table 2** and labelled by following the rule that *u* and *v* refer to the $3d_{3/2}$ and $3d_{5/2}$ spin–orbit components, respectively. In **Table 2** the binding energies and relative peak areas are also reported.

By comparing the *Ce 3d* (and *O 1s*) spectra of the two cases the presence of a shift of one signal respect to the other is immediately evident and reveals the different chemical environment in the two samples, since the spectra have been aligned to *C 1s* peak at 284.6 eV. From the area of the different components of the *Ce 3d* spectrum (**Table 2**) we estimate the $Ce^{3+}$ and $Ce^{4+}$ relative concentration [36], resulting in a $Ce^{3+}$ concentration of 21.8% in $CeO_2$/Si, and 18.0% in $CeO_2$/TiN. Similarly, considering O 1s region, 2 components are identified and, from low to high binding energy (B.E.), assigned to oxygen bonded to Ce and adsorbed – $OH/CO_3^{2-}$ complexes from the environment [18,45]. Since it is well-known that $Ce^{3+}$ oxidation state is preferentially located at the surface of the oxide film [42] which is attributed to the interaction with organics in the air, we can assume that the $Ce^{3+}$ percentage represents the upper limit of the off-stoichiometry for $CeO_2$ layers in both cases. Considering that the O/Ce ratio stoichiometry in a fully oxidized film is 2 for $CeO_2$ or 3/2 for $Ce_2O_3$ and considering the calculated concentrations for $Ce^{3+}$ and $Ce^{4+}$, we obtain an O/Ce ratio of 1.89 for $CeO_2$/Si and 1.91 for $CeO_2$/TiN. Such values indicate a ~10% off-stoichiometry and suggest that $Ce^{3+}$ is possibly related with the presence of oxygen vacancies in $CeO_2$ film, rather than $Ce_2O_3$ amorphous phase. Nonetheless, the difference of $Ce^{3+}$ amount between $CeO_2$/Si and $CeO_2$/TiN, despite small, could be sufficient for inducing a different preferential crystallization in ceria films, as observed by XRD measurement.

It has been reported in literature that the higher is the $Ce^{3+}$ content, the lower is the energy gap value [46]. The optical gap can be estimated by using ellipsometric data. In our case, and in perspective of $CeO_2$ integration in some devices, is relevant to understand if the $Ce^{3+}$ amount and, more generally, the observed structural characteristics and chemical composition of $CeO_2$ films are sensible to a variation of the optical gap [47].

In **figure 5(a)** we report the raw ellipsometric data Ψ (open squares) and Δ (circles) of $CeO_2$/Si. The continuous lines superimposed to the data are the fitting curves obtained from two Tauc-Lorentz (2TL) dispersion model [48]. In **Table 3** the fit parameter values for both $CeO_2$/Si and $CeO_2/TiN$ are summarized. We used a simple three-layer model consisting of $CeO_2$/$SiO_2$-interfacial layer/Si film for $CeO_2$/Si, and a model consisting of $CeO_2$/TiN-oxidized/TiN/ $SiO_2$/Si for $CeO_2$/TiN. The model parameters for TiN layer were firstly determined by a preliminary measurement of such layer before their inclusion in the model for $CeO_2$/TiN.

Keeping $CeO_2$ layer thickness set as variable, the goodness of the chosen model is the correspondence of this thickness value with the thickness obtained by XRR. The obtained value of 24.5 nm (see **Table 3**) for $CeO_2$ on Si, in line with the XRR thickness (considering that the error in the XRR thickness equals to the surface roughness), corroborates the choice of two Tauc-Lorentz oscillators for optical data parametrization.

From the model, we extract the real ($n$) and imaginary ($k$) part of the refractive index in the range 1.24-5.05 eV, showed in **figure 5(b)** for $CeO_2$/Si. In particular, the $n$ value is 2.3 at 2.25 eV (i.e. 550 nm of wavelength), which corresponds to the value estimated for plasma enhanced ALD-deposited $CeO_2$ film [16]. This value is in line with the refractive index found for $CeO_2$ deposited with other techniques [41], but lower than the value measured for highly oriented $CeO_2$ [49]. For $CeO_2$/TiN we obtain $n = 2.4$ at 2.25 eV, close to the value extracted for $CeO_2$ on Si.

From the extinction coefficient $k$, the plot $(\alpha h\nu)^2$ vs $h\nu$ e $(\alpha h\nu)^{1/2}$ vs $h\nu$ can be drawn, **figures 6(a)-(b)**. The optical band gap for the direct and indirect transitions are calculated as the intercept position of the extrapolated linear part of the curves. The obtained optical band gap for the direct transition is $3.54\pm0.63$ eV and for the indirect transition is $2.89\pm0.19$ eV for $CeO_2$/Si, **figure 6(a)**, and $3.63\pm0.18$ eV and $3.05\pm0.22$ eV for the direct and indirect transition for $CeO_2$/TiN, **figure 6(b)**. These values are in good agreement with literature [16, 46, 47]. Our findings suggest that the optical properties of $CeO_2$ are not affected by the differences arisen from the detailed structural and chemical analysis of the films.

**Conclusions**

In summary, in this work we report about a detailed structural and chemical characterization of $CeO_2$ thin films deposited by ALD on both Si and TiN substrates. The ALD growth of the cerium oxide resulted in a polycrystalline structure as expected from previous reports, however with a preferred orientation of the grains depending on the substrate. In particular, we observe an enhancement of the diffraction signal connected to the <100> orientation of the grains in the case of Si substrate, compared to the enhancement of the diffraction signal related to the <111> direction for the TiN substrate. The not-favoured exposure of *(100)* planes for strongly ionic oxide like $CeO_2$ can be explained considering the higher interface $CeO_2$/TiN roughness compared to $CeO_2$/Si one, which can contribute in a change to the energy requirement for the nucleation of crystals with *(111)* orientation. Additionally, the presence of defects or charge compensating species, i.e. oxygen vacancy due to $Ce^{3+}$: $Ce^{4+}$ substitution, can contribute to the balance of the energy requirement. As a perspective for future work, it could be of interest to determine the stoichiometry of the $CeO_2$ in the first layers in contact with the substrate, in order to clarify the role of the interface in dictating the structural properties of the oxide.

However, our study evidences that the substrate plays a role in the growth of $CeO_2$ films by ALD, contributing to boost the consideration for ALD based processes targeting a wide range of applications and a viable route for ceria oxide engineering to address optimal properties for specific applications, such as in microelectronics and catalysis.

AUTHOR INFORMATION

**Corresponding Author**

*S. Vangelista, silvia.vangelista@mdm.imm.cnr.it

**Author Contributions**

The manuscript was written through contributions of all authors. All authors have given approval to the final version of the manuscript.

**Funding Sources**

This work was partially supported by ECSEL-JU R2POWER300 project under grant agreement n.653933.

**ACKNOWLEDGMENTS**

The authors would thank Dr. Sabina Spiga for fruitful discussions, Dr. Elena Cianci (IMM-CNR) for discussions on SE measurements and modelling, Daniela Brazzelli, Fabrizio Toia and Donata Piccolo (ST Agrate) for discussions on sample preparation and Davide Codegoni (ST Physics Lab Agrate) for TEM data and discussion.12

**FIGURES CAPTIONS**

**Figure 1.** XRR experimental curves and data fitting of the as-deposited $CeO_2$ on Si **(a)** and on TiN **(b)** substrates; the corresponding AFM image of the sample is shown as inset.

**Figure 2.** GIXRD experimental pattern and data fitting of as-deposited $CeO_2$ on Si (light blue circles and continuous blue line) and on TiN (black squares and continuous red line) substrates. The pink line is the diffraction powder pattern from cubic (*Fm-3m*) $CeO_2$ used as reference [35].

**Figure 3.** ToF-SIMS depth profile of $CeO_2$ and corresponding EDX maps; **(a)** on Si and **(b)** on TiN substrate.

**Figure 4. (a)** *Ce 3d* XPS spectra (both experimental and fitting curves) collected at take-off angle of 45° from $CeO_2$ on Si (upper panel) and on TiN (lower panel). The fit components of the $Ce^{3+}$ and $Ce^{4+}$ – related doublets are also shown; **(b)** *O 1s* XPS spectra (experimental and fitting curves) from $CeO_2$ on Si (upper panel) and on TiN (lower panel). The 2 components of the profile deconvolution are also shown.

**Figure 5. (a)** Spectroscopic ellipsometric data (open circles and squares) obtained from $CeO_2$ on Si. The full lines are the fit from the dispersion model (2TL); **(b)** real (*n*) and imaginary (*k*) part of the refractive index obtained from the model.

**Figure 6.** Plot of $(\alpha h\nu)^2$ as a function of photon energy (blue open squares); from the linear part of the plot (light blue continuous line) we obtain the direct optical band gap. Plot of $(\alpha h\nu)^{1/2}$ as a function of photon energy (black open triangles); from the linear part of the plot (red continuous line) the indirect optical band gap is obtained. (a) $CeO_2$/Si; (b) $CeO_2$/TiN.

# TABLES

**Table 1.** Properties of $CeO_2$ grown on Si and on TiN from XRR and XRD data fitting.

| Sample ID | Thickness (nm) | Interface roughness (nm) | Surface roughness (nm) | El. Density (e/Å$^3$) | Grain size (nm) (111) | Grain size (nm) (200) |
|---|---|---|---|---|---|---|
| $CeO_2$ on Si | 23.8±0.3 | 0.3±0.1 | 1.3±0.1 | 1.69±0.05 | 8.0±0.1 | 11.6±0.1 |
| $CeO_2$ on TiN | 23.8±0.2 | 0.8±0.1 | 1.4±0.1 | 1.65±0.05 | 8.3±0.1 | 11.4±0.1 |

**Table 2.** XPS peak assignment, positions and relative area of $Ce_{3d}$ peaks. The areas of the different components are used to calculate the percentage of $Ce^{3+}$ and $Ce^{4+}$.

| Sample ID | Peak assignment | Ce species | Binding energy (eV) | Relative area (%) |
|---|---|---|---|---|
| $CeO_2$ on Si | v$_0$ | $Ce^{3+}$ | 882.0 | 9 |
| | v | $Ce^{4+}$ | 882.4 | 24 |
| | v' | $Ce^{3+}$ | 885.3 | 6 |
| | v'' | $Ce^{4+}$ | 888.5 | 10 |
| | v''' | $Ce^{4+}$ | 898.1 | 18 |
| | u$_0$ | $Ce^{3+}$ | 900.3 | 4 |
| | u | $Ce^{4+}$ | 900.7 | 12 |
| | u' | $Ce^{3+}$ | 903.6 | 3 |
| | u'' | $Ce^{4+}$ | 906.8 | 5 |
| | u''' | $Ce^{4+}$ | 916.3 | 9 |
| | **Percentage of $Ce_2O_3$ = 22 %** | | | |
| | **Percentage of $CeO_2$ = 78 %** | | | |
| $CeO_2$ on TiN | v$_0$ | $Ce^{3+}$ | 881.9 | 7 |
| | v | $Ce^{4+}$ | 882.4 | 21 |
| | v' | $Ce^{3+}$ | 885.1 | 5 |
| | v'' | $Ce^{4+}$ | 888.5 | 12 |
| | v''' | $Ce^{4+}$ | 897.9 | 21 |
| | u$_0$ | $Ce^{3+}$ | 900.2 | 3 |
| | u | $Ce^{4+}$ | 900.7 | 11 |
| | u' | $Ce^{3+}$ | 903.4 | 3 |
| | u'' | $Ce^{4+}$ | 906.8 | 6 |
| | u''' | $Ce^{4+}$ | 916.3 | 11 |
| | **Percentage of $Ce_2O_3$ = 18 %** | | | |
| | **Percentage of $CeO_2$ = 82 %** | | | |

**Table 3.** 2TL model parameters for the $CeO_2$ film on Si and TiN. $\varepsilon\infty$ is the value of real part of the dielectric function $\varepsilon$ at infinite energy, Eg is the bandgap of the material, A is the oscillator amplitude, E0 is the energy (position) of the Lorentz peak, and C is the broadening parameter.

| Sample ID | Thickness $CeO_2$ (nm) | $\varepsilon\infty$ (eV) | A1 | Eg1 (eV) | E01 (eV) | C1 | A2 | Eg2 (eV) | E02 (eV) | C1 |
|---|---|---|---|---|---|---|---|---|---|---|
| $CeO_2$ on Si | 24.5±0.01 | 2.4±0.1 | 51.6±1.1 | 2.8±0.01 | 3.9±0.01 | 0.8±0.01 | 41.8±4.5 | 2.9±0.01 | 6.4±0.1 | 4.5±0.7 |
| $CeO_2$ on TiN | 25.0±0.01 | 1.7±0.1 | 87.1±2.0 | 3.0±0.01 | 3.8±0.01 | 0.8±0.01 | 31.1±3.5 | 3.1±0.01 | 7.0±0.1 | 2.9±0.5 |

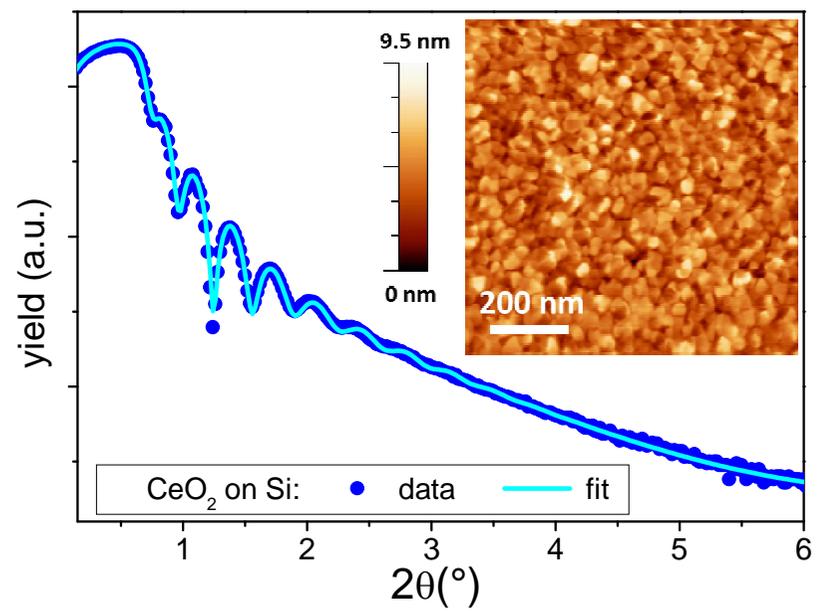 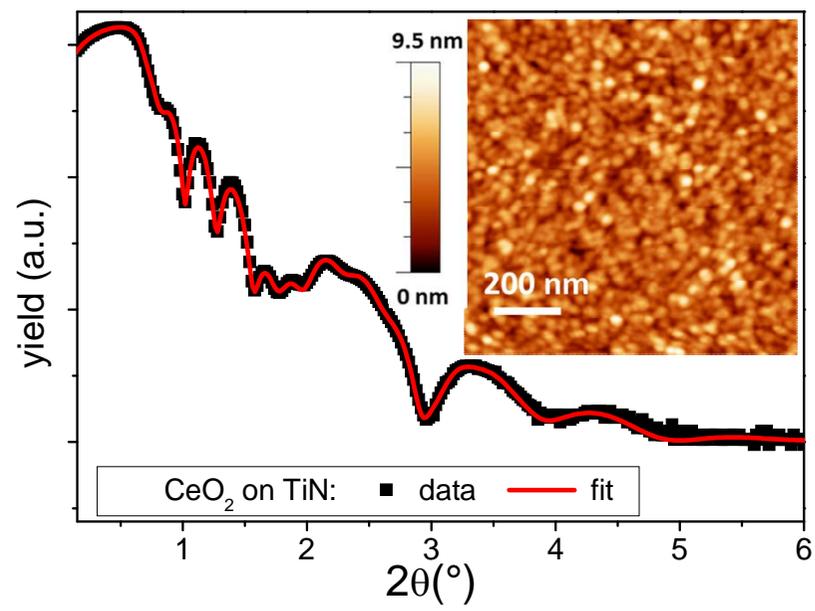

**(a)** **(b)**

**Figure 1**

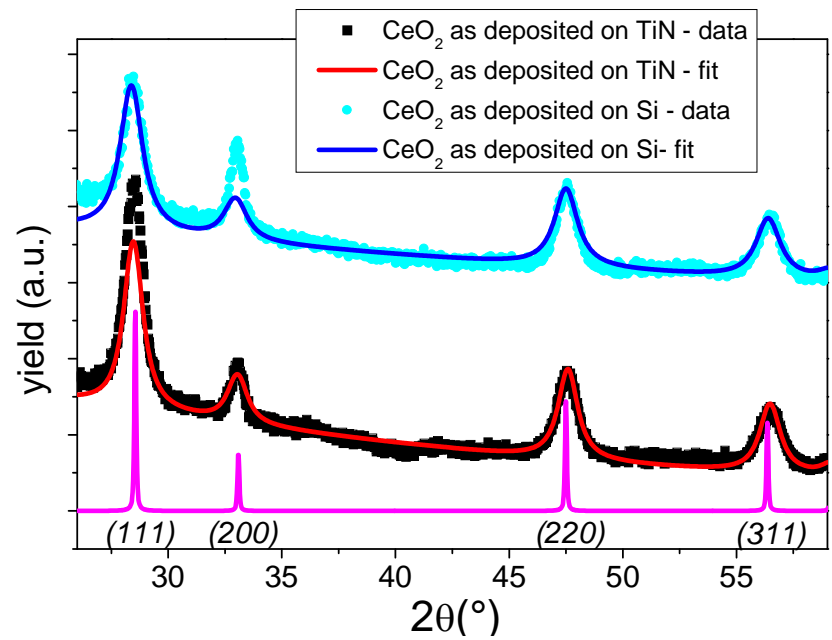

**Figure 2**

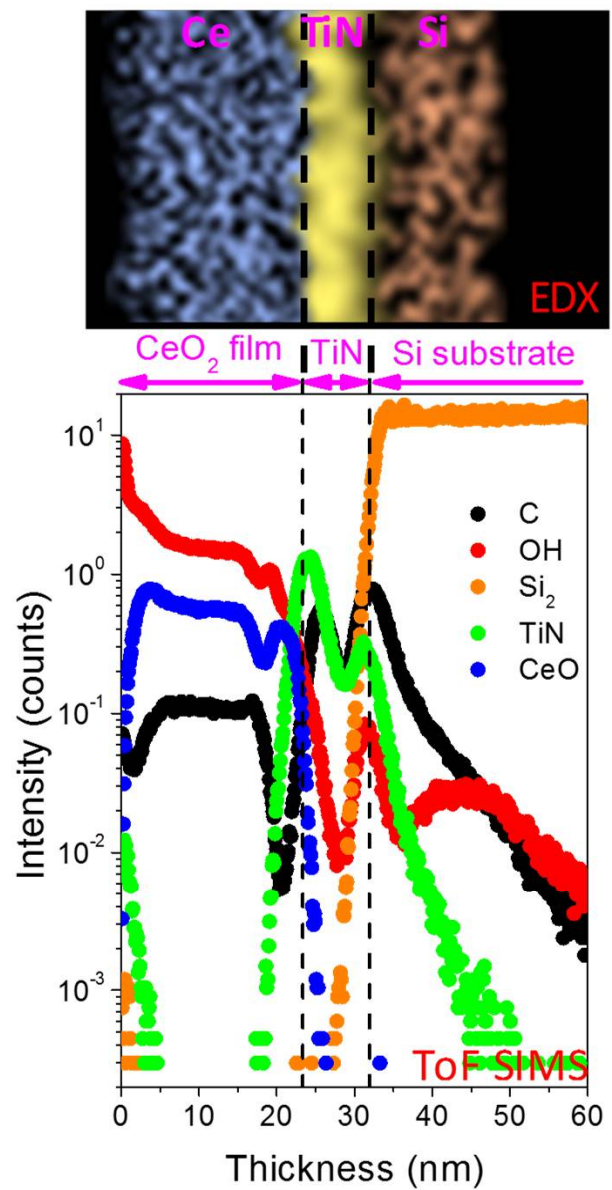 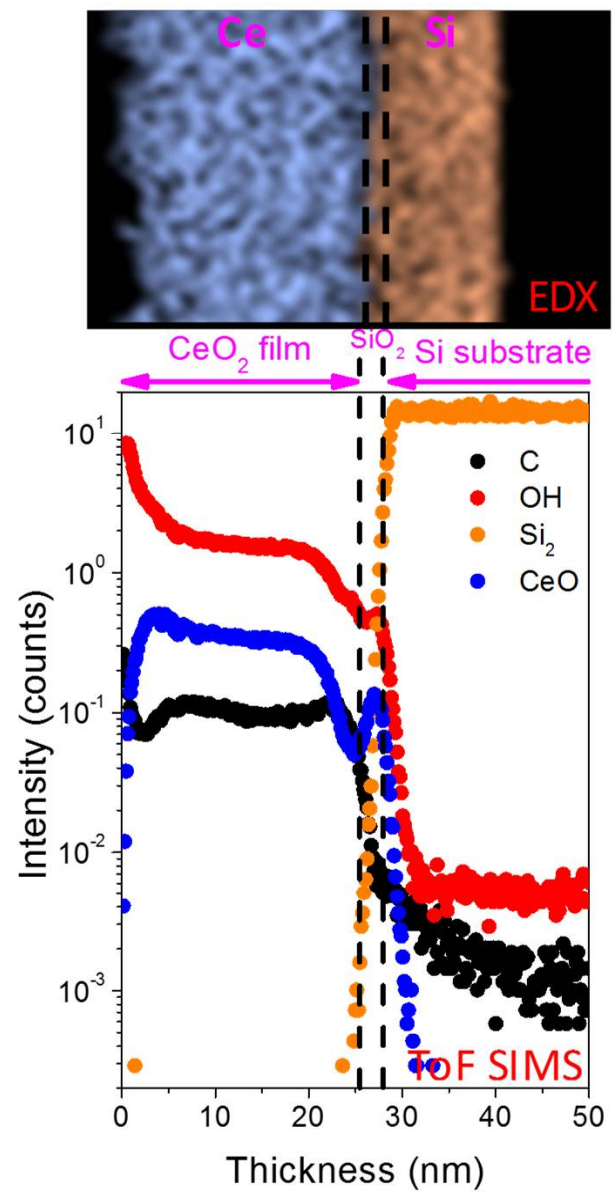

(a)                          (b)

**Figure 3**

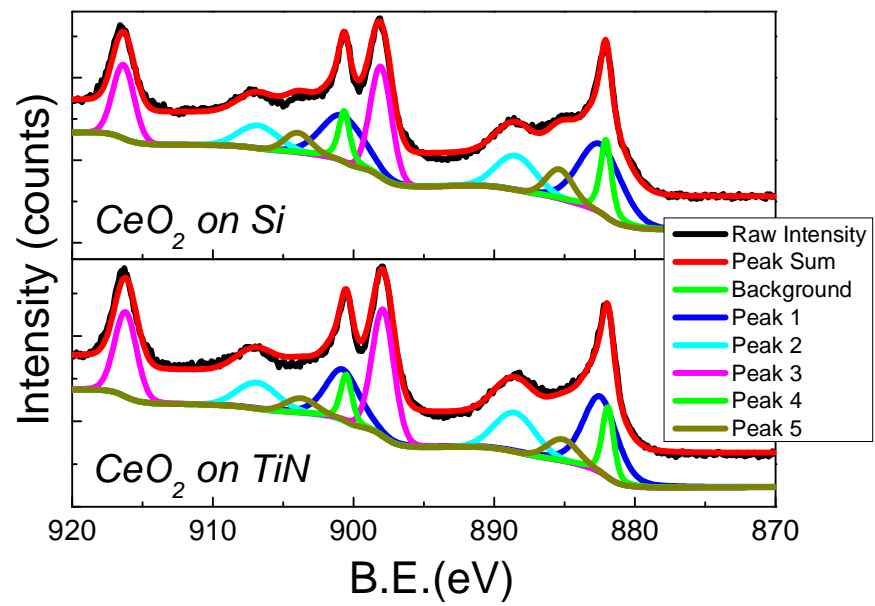
(a)

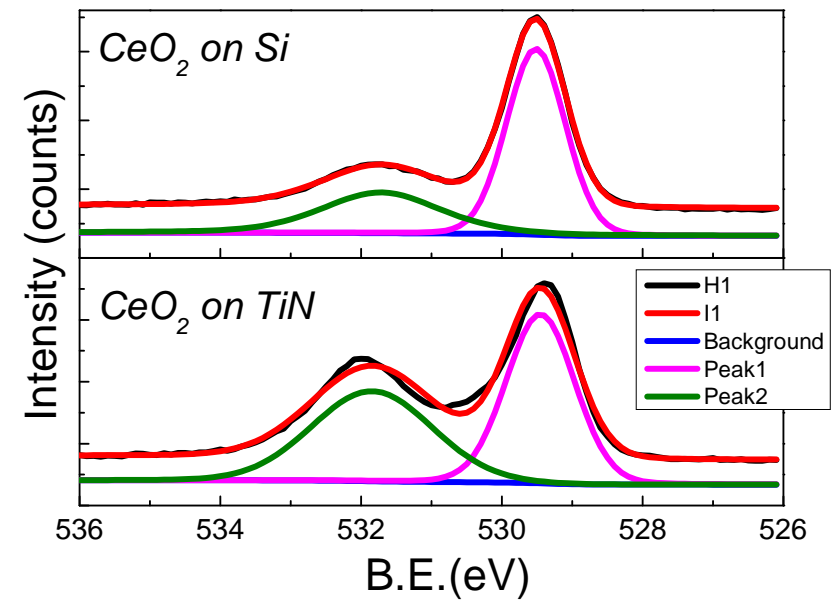
(b)

**Figure 4**

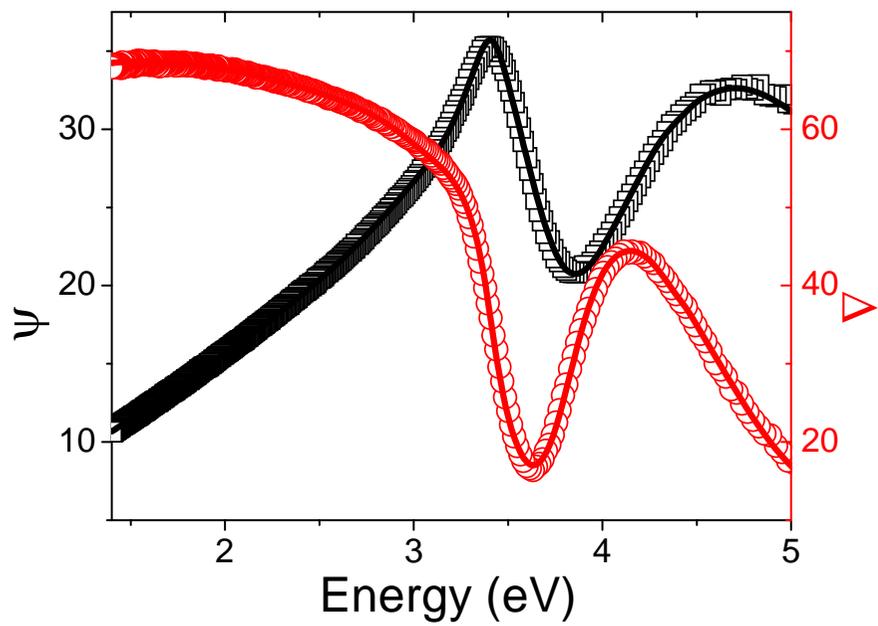 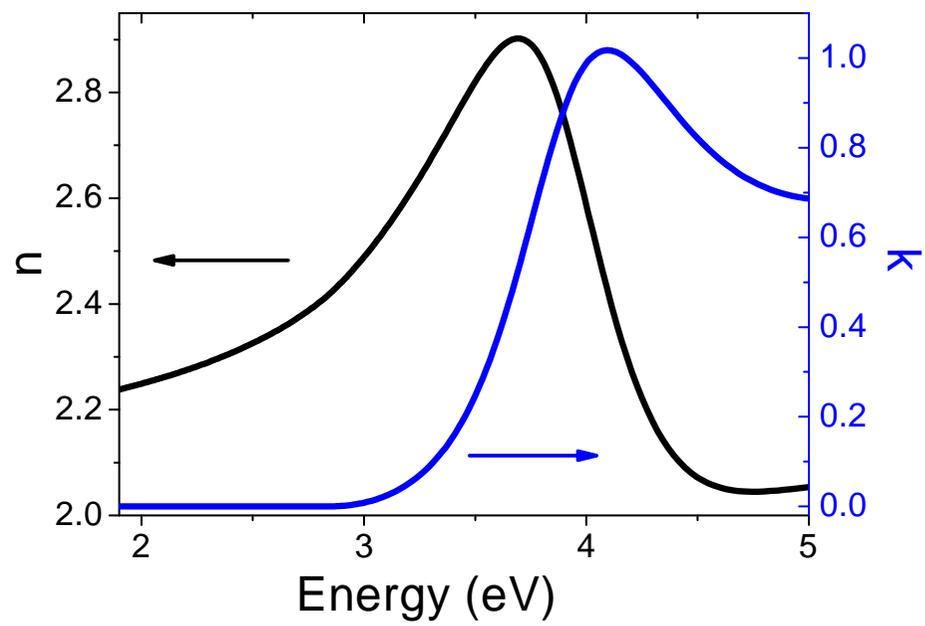

(a) (b)

**Figure 5**

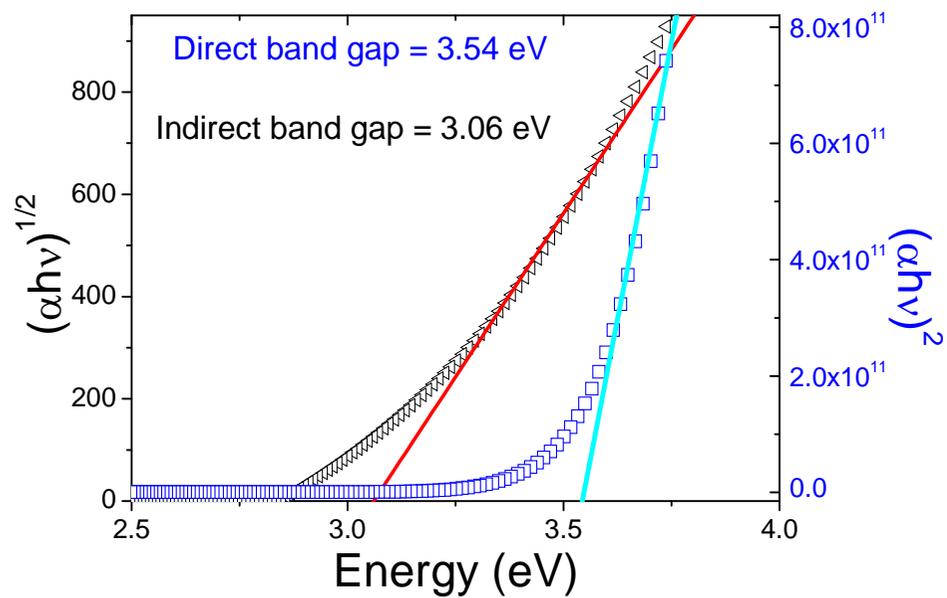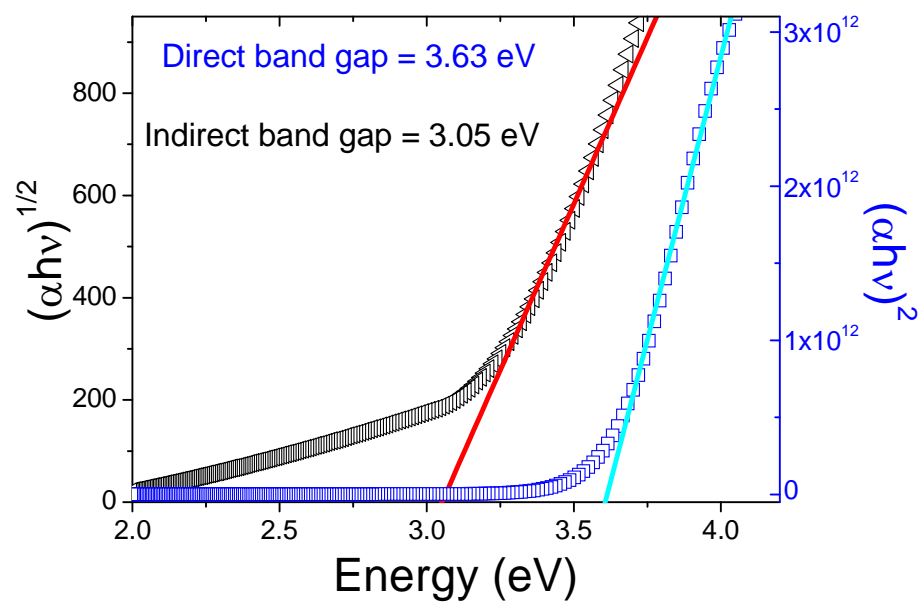

(a) (b)

**Figure 6**